\documentstyle{article}
\begin{document}
\baselineskip 18pt
\title{The general crossing relation for boundary reflection matrix }
\author{Bo-yu Hou$^{\ddagger}$ \hskip 0.5truecm , Kang-jie Shi$^{\dagger,
          \ddagger}$\\
              and Wen-li Yang$^{\dagger,\ddagger}$\\
\bigskip\\
$^{\dagger}$ CCAST (World Laboratory),PO Box 8730,Beijing 100080,China
\\
$^{\ddagger}$ Institute of Modern Physics,Northwest University,Xian 710069,
China
\thanks{Mailing address}}
\maketitle
\begin{abstract}
{In this paper, we give the general crossing relation for boundary reflection 
matrix $R(\beta)$  ,which is the extension of the work given by 
Ghoshal and Zamolodchikov .We also use the first non-trivial extended 
crossing relation to determine the scaler factor of $R(\beta)$ which is the 
rational diagonal solution to the boundary Yang-Baxter equation in the case 
of l=2 and n=3.}
\end{abstract}

\newpage

\section{Introduction}
Much work has recently been done in integrable quantum field theory and 
lattice statistical mechanics on models with a boundary , where the 
integrability is guaranteed by the boundary Yang-Baxter equation ( BYBE ) .
An exact solution to such a field theory could provide a better undstanding 
of boundary-related phenomena in statistical systems near criticality$^{[1]}$.
In statistical mechanics,the emphasis has been on deriving solutions of BYBE 
and calculation of various surface critical phenomena, both at and away from  
criticality$^{[8,9]}$.

An important character of an integrable field theory is its 
factorizable scattering matrix (S-matrix). In the ``bulk theory" ,
the factorizable S-matrix is completely determined in terms of the 
two-particle S-matrix ,which should satisfy the Yang-Baxter equation(YBE), 
in addtion to the standard equations of unitarity and crossing relation$^{[2,3]}$ .
The general crossing relation was given by Y.H. Quano$^{[6]}$ .These equations 
have much restrictive power, determining the S-matrix up to the so-called ``CDD 
ambiguity" .Recently, much important progress in the boundary integrable 
model is that Ghoshal and Zamolodchikov gave the approprite anolog of 
the crossing relation for boundary reflection matrix (R-matrix) in the Ref.[3].Together with BYBE and 
unitarity  for boundary R-matrix,the above equations have exactly the same 
restrictive power as the corresponding ``bulk " system ,i.e. they allow one 
to pin down the facterizable boundary R-matrix up to the ``CDD factors" .
This crossing relation also play a very important role in solving boundary 
integrable spin chain (e.g boundary XXZ model $^{[4]}$ and boudary 
XYZ model$^{[5]}$). However, it was known that the crossing relation gave by 
Ghoshal and Zamolodchikov is only the special case (n=2) of the general one.
In this paper, we will give the general version of crossing relation for 
boundary R-matrix (i.e for any $A_{n}^{(1)}$ case)

\section{The general boundary crossing relation}
For an integrable boundary massive field theory, the scattering theory is 
purely elastic and the corresponding scattering matrix is factorizable. 
Besides the bulk scattering processes (which is described by two-particle 
S-matrix $S(\beta)$),there exist a process which represents the particles 
reflecting with the boundary.This process is described by boundary 
R-matrix $R(\beta)$ .The 
factorization of scattering in the boundary case is equivalent to validity 
of YBE and BYBE
\begin{eqnarray}
& &{\rm YBE} \ \ :\ \ \  S_{12}(\beta_{1}-\beta_{2})S_{13}(\beta_{1}-
\beta_{3})S_{23}(\beta_{2}-\beta_{3})\nonumber\\
& &\ \ \ \ \ \ \ =S_{23}(\beta_{2}-\beta_{3})
S_{13}(\beta_{1}-\beta_{3})S_{12}(\beta_{1}-\beta_{2})\ \ \ ,\\
& &{\rm BYBE }\ \ :  \ \ S_{12}(\beta_{1}-\beta_{2})R_{1}(\beta_{1})
S_{21}(\beta_{1}+\beta_{2})R_{2}(\beta_{2})\nonumber\\
& &\ \ \ \ \ \ \ =R_{2}(\beta_{2})S_{12}(\beta_{1}+\beta_{2})R_{1}(\beta_{1})
S_{21}(\beta_{1}-\beta_{2})\ \ \ \ ,
\end{eqnarray}
where $S(\beta)$ is the same one as in the bulk case 
 and  boundary R-matrix should satisfy the requirements of unitary  
\begin{eqnarray}
R(\beta)R(-\beta)=1\ \ \ \ .
\end{eqnarray}
AS the usual conventions,$S_{mn}(\beta)$ signifies the matrix on $V\otimes V
...\otimes V$ acting on the m-th and n-th component and as identity on the 
other ones, and $R_{k}(\beta)$ signifies the matrix on tensor space acting
on th k-th component and as identity on the other ones.

\subsection{The general boundary crossing relation}
We only consider the solution to YBE which has the following property
\begin{eqnarray}
S(-w)=P^{(-)}_{2}\times ({\rm something)}\ \ \ \ , \label{project}
\end{eqnarray}
where $ w$ is so-called crossing parameter (it is usally rescaling as $i\pi$ )
,and $P^{(-)}_{2}$ is the completely 
antisymmetric project operator in $V^{\otimes 2}$. In fact, in addition to 
rational and elliptic solution to YBE having properties (~\ref{project})
,there exist a
series trigonometric solution to YBE (which is the trigonometrical limit of 
Zn Belavin' solution ) carrying with property (~\ref{project}).

Due to the property (~\ref{project}),one can construct the completely 
antisymmetric
fusion procedure for R-matrix$^{[10,11,12]}$.First, we give some notation about 
fusion.
Let $V=C^{n}$ ( 2$\leq$ n) and $V^{*}$ be the dual space of $V$ .Then we 
have a homorphism $C$ :
\begin{eqnarray*}
C:\ \ \ V^{*}\longrightarrow \Lambda ^{n-1}(V) \ \ \ {\rm and}\\
Ce^{*}_{i}=\frac{1}{\sqrt{(n-1)!}}\epsilon^{i_{1}...i_{n-1}}_{i}
e_{i_{1}}\otimes...\otimes e_{i_{n-1}}\ \ \ \ \ ,
\end{eqnarray*}
where $e^{*}_{i}$ is the base of dual space $V^{*}$ and $\epsilon^
{i_{1}....i_{n-1}}_{i}$ is the n-th order completely antisymmetric tensor.
Through the fusion procedure for R-matrix ,one can obtain 

\noindent {\large \bf Theorem 2.1 } For any integral $l\ \ (l\leq n)$ , 
define 
\begin{eqnarray*}
R^{(l)}(\beta) :\ \ \ \ \ V^{\otimes l}\longrightarrow V^{\otimes l} \ \ \ 
{\rm by }\\
R^{(l)}(\beta)=R_{l}(\beta +lw)S_{l-1,l}(2\beta +(2l-1)w)...R_{1}(\beta+w)
\end{eqnarray*}
then image of $R^{(l)}(\beta)|_{\Lambda^{(l)}(V)}$ is also in 
$\Lambda^{(l)}(V)$ .Furthermore,we can define 
\begin{eqnarray*}
R^{*}(\beta):\ \ \ \ \ V^{*}\longrightarrow V^{*}\ \ \ \ {\rm by}
\end{eqnarray*}
\begin{eqnarray}
& &R^{*}(\beta)=C^{-1}P^{(-)}_{n-1}R_{n}(\beta+(n-1)w)S_{n-1,n}
(2\beta+(2n-3)w)\nonumber\\
& &\ \ \ \ .....R_{2}(\beta +w)P^{(-)}_{n-1}C\ \ \ \ \ .
\end{eqnarray}

\noindent [{\large \bf Proof :}] 

From YBE and the properties (~\ref{project}), it is easy to see the 
following properties
\begin{eqnarray}
& &S_{12}(-w)S_{13}(-2w)...S_{1l}(-(l-1)w)S_{23}(-w)....S_{l-1,l}(-w)
\nonumber\\
& & \ \ \ \ \ \ =P^{(-)}_{l}\times M \ \ \ \ ,\\
& &S_{l,l-1}(-w)S_{l,l-2}(-2w)...S_{l1}(-(l-1)w)S_{l-1,l-2}(-w)....S_{21}
(-w)\nonumber\\
& &\ \ \ \ \ \ \ =P^{(-)}_{l}\times M'\ \ \ \ \ ,
\end{eqnarray}
where $P^{(-)}_{l}$ is completely antisymmetric project operator in 
$V^{\otimes l}$ and $M,M' \in {\rm End}(V^{\otimes l})$ .Using the YBE and 
BYBE, one have
\begin{eqnarray} 
& &[S_{12}(-w)S_{13}(-2w)...S_{1l}(-(l-1)w)S_{23}(-w)....S_{l-1,l}(-w)]\times
\nonumber\\
& & \ \ \ R_{1}(\beta+w)S_{21}(2\beta+3w)...S_{l1}(2\beta+(l+1)w)R_{2}
(\beta+2w)\nonumber\\
& & \ \ \ \ \ ...R_{l}(\beta+lw)\nonumber\\
& &=R^{(l)}(\beta)[S_{l,l-1}(-w)S_{l,l-2}(-2w)...S_{l1}(-(l-1)w)S_{l-1,l-2}
(-w)\nonumber\\
& & \ \ \ \ \ ....S_{21}(-w)]\ \ \ \ \ \ .
\end{eqnarray}
Thanks to properties (6) and (7), this means that
\begin{eqnarray}
P^{(-)}_{l}R^{(l)}(\beta)P^{(-)}_{l}=R^{(l)}(\beta)P^{(-)}_{l}\ \ \ \ ,
\end{eqnarray}
so it ensure us that Im($R^{(l)}(\beta))|_{\Lambda^{(l)}(V)} \in
\Lambda^{(l)}(V)$ .

\noindent We are now in a position to mention the crossing relation for 
generic n .Because the crossing relation of boundary R-matrix is mainly 
considered,let us assume that our $R(\beta)$ has already enjoyed in the 
unitarity :$R(\beta)R(-\beta)=1$ , and $S(\beta)$ enjoyed in unitarity 
and crossing symmetry$^{[6]}$
\begin{eqnarray}
& &S_{12}(\beta)S_{21}(-\beta)=1\ \ \ \ ,\nonumber\\
& &[S^{V,V^{*}}(\beta)]^{jl^{*}}_{ik^{*}}=S(-\beta-nw)^{kj}_{li}\ \ \ \  ,
\end{eqnarray}
where 
\begin{eqnarray*}
& &S^{V,V^{*}}(\beta)=(1\otimes C^{-1})(1\otimes P^{(-)}_{n-1})
S_{1n}(\beta+(n-1)w)....S_{12}(\beta+w)\\
& &\ \ \ \ \ (1\otimes P^{(-)}_{n-1})(1\otimes C)\ \ \ \ \ .
\end{eqnarray*}

Using the definition of $R^{*}(\beta)$ ,the unitarity  
of $S(\beta)$ and  $R(\beta)$ ,one can find $R^{*}(\beta)$ with the
following inversion relation
\begin{eqnarray}
R^{*}(\beta)R^{*}(-\beta-nw)=1\ \ \ \ .
\end{eqnarray} 

\noindent {\large \bf Theorem 2.2} $R^{*}(\beta)$ which is obtained through 
fusion of $R(\beta)$ can also be directly expressed into 
\begin{eqnarray}
[R^{*}(\beta)]^{k}_{j}=D(\beta)[S^{V,V}(2\beta+nw)]^{lj}_{kl'}[R(-\beta-nw)]
^{l'}_{l}\ \ \ \ ,
\end{eqnarray}
where $D(\beta)$ is a scaler factor (actually it is the quantum determinator 
of $R(\beta)$ )

\noindent [{\large \bf Proof:}] 

Owing to the fact that dim($\Lambda^{n}(V)$ is equal 
to 1 and Theorem 2.1 ,one have
\begin{eqnarray*}
\sum _{m,l,l'}[R^{*}(\beta)]^{k}_{m}[S^{V,V^{*}}(2\beta)]^{jm}_{l'l}
[R(\beta)]^{l'}_{l}=D(\beta)\delta^{jk}\ \ \ \ ,
\end{eqnarray*}
where $D(\beta)$ is a scaler factor which depends on $R(\beta)$ and 
$S(\beta)$ .Using the inversion relation (11) and the crossing symmetry of
$S(\beta)$ (10), one can obtain
$$
[R^{*}(\beta)]^{k}_{j}=D(\beta)[S^{V,V}(2\beta+nw)]^{lj}_{kl'}[R(-\beta-nw)]
^{l'}_{l}\ \ \ \ \ .\eqno(12a)
$$

Due to (12),one can reduce $D(\beta)$ to 1 through rescaling the boundary 
matrix $R(\beta)$ .Therefore,one can determine the scaler factor of 
$R(\beta)$ using the following relation
\begin{eqnarray*}
[R^{*}(\beta)]^{k}_{j}=[S^{V,V}(2\beta+nw)]^{lj}_{kl'}[R(-\beta-nw)]
^{l'}_{l}\ \ \ .
\end{eqnarray*}
This relation can be considered as the generalization crossing relation for 
generic n.When one considers $V^{*}$ ,circumstances slightly change between 
n=2 and $n>2$ case . Speaking in terms of Young tableau , $V=\Box$ ,the
fundamental representation $sl_{n}$, while $V^{*}$ corresponds to 
vertical n-1 $\Box $'s ,the space of antisymmetric tensors in 
$V^{\otimes n-1}$ .Thus $V$ and $V^{*}$ can be identified even at the level 
of Young tableau if and only if n=2 .For $n>2$ , more complex procedure than
(5) is needed and (12a) is non-trivial extension of crossing relation for
$n>2$ .
However,when n is equal to 2 ,the restricted condition for scaler factor of 
$R(\beta)$ become the same as that given by Ghoshal and Zamolodchikov.We 
have checked it for the rational ,trigonometric and elliptical type solution 
to BYBE.

Hence,in order to determine the exact boundary reflecting matrix ,one can 
impose the following condition on  $R(\beta)$ (BYBE,unitarity,crossing 
relation)  
\begin{eqnarray}
& &{\rm BYBE \ \ :} \ \ S_{12}(\beta_{1}-\beta_{2})R_{1}(\beta_{1})S_{21}
(\beta_{1}+\beta_{2})R_{2}(\beta_{2})\nonumber \\
& &\ \ \ \ \ \ \ =R_{2}(\beta_{2})S_{12}(\beta_{1}+\beta_{2})R_{1}(\beta_{1})
S_{21}(\beta_{1}-\beta_{2})\ \ \ \ \ ,\\
& &{\rm unitarity\ \:} \ \ \ \ \ R(\beta)R(-\beta)=1\ \ \ \ ,\\
& &{\rm crossing \ \ relation \ \ :}\ \ [R^{*}(\beta)]^{k}_{j}=[S^{V,V}(2\beta+nw)
]^{lj}_{kl'}[R(-\beta-nw)]^{l'}_{l}\ \ \ \ .
\end{eqnarray}
The above condition can only reduce the scaler factor up to some 
``CDD" factor $\Phi (\beta)$ which should also be generalized as satisfying 
follow requirement
\begin{eqnarray}
\Phi (\beta)\Phi (-\beta)=1\ \ ,\ \ \prod^{n}_{l=1} \Phi (\beta +lw)=1\ \ \ .
\end{eqnarray}
When n=2 ,the above requirement return the usual one.Moreover,this ambiguity 
can be canceled by some dynamical requirement$^{[2,3]}$.
\subsection{An example}
To determine the exact boundary reflecting matrix $R(\beta)$ ,the following 
procedure are usually taken : first,one obtain the solution to (13) upon to
some scaler factor;then, the unitarity (14) and crossing relation (15) are
used to determine the scaler factor.In this subsection, we obtain the 
rational diagonal solution to BYBE for generic n. Then without losing the 
generality, the first nontrivial extension of crossing relation are applied 
to determine the scaler factor for n=3.

Here,we consider the rational solution to YBE with unitarity and crossing 
relation satisfied,which is given as follows
\begin{eqnarray*}
& &S(\beta)=a(\beta)1+b(\beta)P\\   
& &\ \ \ =k(\beta)(\frac{\beta}{\beta+w}1+\frac{w}{\beta+w}P)\ \ \ ,
\end{eqnarray*}
where P is the permutation operator in $V^{\otimes 2}$:
\begin{eqnarray*}
P(e_{i}\otimes e_{j})=e_{j}\otimes e_{i}
\end{eqnarray*}
and scaler factor $k(\beta)$ is equal to 
\begin{eqnarray*}
k(\beta)=-\frac{\Gamma(\frac{\beta}{nw})\Gamma(\frac{-\beta}{nw}+
\frac{1}{nw})} {\Gamma(\frac{-\beta}{nw})\Gamma(\frac{\beta}{nw}+
\frac{1}{nw})}\ \ \ \ .
\end{eqnarray*}
The BYBE leads to the following functional equations
\begin{eqnarray*}
& &a(\beta_{1}+\beta_{2})b(\beta_{1}-\beta_{2})R^{k}_{j}(\beta_{1})R^{l}_{i}
(\beta_{2})+a(\beta_{1}-\beta_{2})b(\beta_{1}+\beta_{2})R^{i'}_{i}(\beta_{1})
R^{l}_{i'}(\beta_{2})\delta^{k}_{j}\\
& &\ \ \ \ \ \ +b(\beta_{1}+\beta_{2})b(\beta_{1}-\beta_{2})R^{i'}_{j}
(\beta_{1})R^{l}_{i'}(\beta_{2})\delta^{k}_{i}\\
& &=a(\beta_{1}+\beta_{2})b(\beta_{1}-\beta_{2})R^{l}_{i}(\beta_{1})R^{k}_{j}
(\beta_{2})+a(\beta_{1}-\beta_{2})b(\beta_{1}+\beta_{2})R^{k}_{i'}(\beta_{1})
R^{i'}_{j}(\beta_{2})\delta^{l}_{i}\\
& &\ \ \ \ \ \ +b(\beta_{1}+\beta_{2})b(\beta_{1}-\beta_{2})R^{l}_{i'}
(\beta_{1})R^{i'}_{j}(\beta_{2})\delta^{k}_{i}\ \ \ \ .
\end{eqnarray*}
In the diagonal case ,i.e $R^{j}_{i}(\beta)=R_{i}(\beta)\delta^{j}_{i}$ , the 
above equations become
\begin{eqnarray}
& &a(\beta_{1}+\beta_{2})b(\beta_{1}-\beta_{2})[R_{j}(\beta_{1})
R_{i}(\beta_{2})-R_{i}(\beta_{1})R_{j}(\beta_{2})]\nonumber\\
& &=a(\beta_{1}-\beta_{2})b(\beta_{1}+\beta_{2})[R_{j}(\beta_{1})
R_{j}(\beta_{2})-R_{i}(\beta_{1})R_{i}(\beta_{2})]\ \ .
\end{eqnarray}
The solution to (17) is
$$
R(\beta)= \left\{
\begin{array}{ll}
r(\beta)\ \ \ ,&if\ \ \ i\leq n-l\\
r(\beta)\frac{\xi-\beta}{\xi+\beta}\ \ ,&otherwise
\end{array}
\right.\eqno(17a)
$$
This diagonal solution has two parameters :$l\in N$ and $\xi \in C$ ,which 
has the same structure as the diagonal solution for trigonometric solution 
given by de Vega$^{[7]}$ .

Only for simplicity but without losing the generality,we only consider the 
crossing relation of this 
solution (17a) for l=2 and n=3,namely,$R(\beta)$ is equal to
\begin{eqnarray}
r(\beta)  \left(
   \begin{array}{llcl}
    1& & &\\
     &1& &\\
     & &\frac{\xi -\beta}{\xi+\beta}
    \end{array}
  \right)\ \ \ \ .
\end{eqnarray}
Using the fusion for $R(\beta)$ ,one can get 
\begin{eqnarray}
& &R^{*}(\beta)=\frac{r(\beta+w)r(\beta+2w)k(2\beta+3w)(\xi -\beta -2w)
(\beta +w)}{(\beta+2w)(\xi +\beta +w)}\nonumber\\
& &\ \ \ 
  \times \left(
   \begin{array}{llcl}
    1& & &\\
     &1& &\\
     & &\frac{\xi +\beta +w}{\xi-\beta -2w}
    \end{array}
  \right)\ \ \ \ .
\end{eqnarray}
On the other hand,from the crossing relation (15), one should obtain
\begin{eqnarray}
& &R^{*}(\beta)=\frac{r(-\beta -3w)k(2\beta+3w)(\beta+3w)(\xi-\beta-2w)}
{(\beta+2w)(\xi-\beta-3w)}\nonumber\\
& &\ \ \ \times \left(
   \begin{array}{llcl}
    1& & &\\
     &1& &\\
     & &\frac{\xi +\beta +w}{\xi-\beta -2w}
    \end{array}
  \right)\ \ \ \ .
\end{eqnarray}              
So, unitarity and crossing relation (14) and (15) lead to the requirements 
for the scaler factor $r(\beta)$
\begin{eqnarray*}
& &r(\beta)r(-\beta)=1 \ \ \ \ ,\\
& &r(\beta+w)r(\beta+2w)r(\beta+3w)=\frac{(\xi+\beta+w)(\beta+3w)}{(\xi-\beta-3w)
(\beta+w)}\ \ \ \ .
\end{eqnarray*}
Solving the above equations,one can obtain the scaler factor $r(\beta)$ upon 
to ``CDD" ambiguity,
\begin{eqnarray}
& &r(\beta)=-\frac{\Gamma(\frac{\beta}{3w})\Gamma(\frac{\xi+\beta}{3w}+
\frac{1}{3})} {\Gamma(-\frac{\beta}{3w})\Gamma(\frac{\xi-\beta}{3w}+
\frac{1}{3})}\nonumber\\
& &\ \ \ \times \frac{\Gamma(-\frac{\beta}{3w}+\frac{1}{3})\Gamma(\frac{\xi-
\beta}{3w})} {\Gamma(\frac{\beta}{3w}+\frac{1}{3})\Gamma(\frac{\xi+\beta}
{3w})}\times \Phi(\beta)\ \ \ \ ,
\end{eqnarray}
where $\Phi(\beta) $ is a ``CDD" factor which satisfies equations
\begin{eqnarray}
& &\Phi (\beta)\Phi (-\beta)=1\ \ \ \ ,\nonumber\\
& &\Phi (\beta+w)\Phi (\beta+2w)\Phi (\beta+3w)=1\ \ \ \ .
\end{eqnarray}
The equations (22) which ``CDD" factors should be satisfied is the generalized
version of n=2 case .

\section*{Acknowledgements}
The author W.L. Yang thank Prof.L. Chao for his fruitful discussions.

\end{document}